# Comparison between the influence of outflows and supermassive binary black holes in active galactic nuclei on the polarization angle profiles

**Đorđe Savić** [1,2]

[1] Astronomical Obsevatory of Belgrade, Volgina 7, 11070 Belgrade, Serbia
[2] Université de Strasbourg, CNRS, Observatoire Astronomique de Strasbourg, UMR 7550, 11 rue de l'Université, F-67000 Strasbourg, France
E-mail: djsavic@aob.rs

## Abstract

Optical polarization signal coming from the innermost part of active galactic nuclei (AGNs) is highly sensitive on the geometry and kinematics of the central engine. Due to the compact size of the AGN central region, which is spatially unresolved with current observing facilities, we rely on spectropolarimetry which can provide us insight in their hidden physics. We model equatorial scattering for various broad line region (BLR) configurations using radiative transfer code *STOKES*. We analyze the polarization position angle ($\varphi$) profiles for four supermassive binary black holes (SMBBHs) models and compare them with the profiles found for a unified model in AGNs with a single supermassive black hole (SMBH) and with notable outflowing velocity component of the BLR.. We find that the $\varphi$ profiles for SMBBHs are axis-symmetric, while the profiles for a single SMBHs are point-symmetric and that there is a clear distinction between the two cases. Our conclusion is that spectropolarimetry might play a key role in the search for the SMBBHs by inspecting the polarization angle profiles.

Keywords: term, term, term

## 1. Introduction

Active Galactic Nuclei (AGN) are contributing to only few percent of galaxies in the Universe. The radiation coming from the nucleus is often surpassing the radiation coming from the rest of the galaxy many times. The vast amount of energy emitted is due to the accretion of gas onto the supermassive black hole (SMBH) [1] with typical mass range from $10^6 - 10^9$ Solar masses [2]. AGNs emit powerful broad spectrum continuum from high energy gamma-rays to low energy radio-waves. In optical domain of the AGN spectrum, for some objects we observe prominent broad and narrow emission lines, while for others, only narrow emission lines are present. It is widely accepted that this dichotomy between the two types of AGNs is explained by the so called "unified model" of AGNs [3, 4]. In this model, for every AGN, the SMBH with an accretion disk is situated in the center and it is surrounded by a dusty torus in the equatorial plane. Due to the orientation of the system, when the line of sight towards the central engine is not obscured by the dusty torus, we can observe both broad and narrow emission lines (Type-1 objects). On the other hand, if the central engine is obscured, only narrow emission lines are visible (Type-2 objects). Spectropolarimetric observations of AGNs played crucial pivot towards the unified model [5]. It was found that Type-1 objects have optical continuum polarization position angle $\varphi$ which is parallel to the axis symmetry of the system arising due to the equatorial scattering in the vicinity of the source [6, 7, 8]. On the contrary, Type-2 objects, have $\varphi$ which is usually perpendicular to the symmetry axis and comes predominantly due to Thomson scattering in the ionized media located in the polar regions. Although broad lines are not visible in the unpolarized spectra of Type-2 objects, they are visible in the polarized light [5].

The broad emission lines originate from the broad line region (BLR) that is directly influenced by the mass of the SMBH. We can expect nearly Keplerian motion of the gas [9] and from the line width, the SMBH mass can be estimated. The BLR gas density is of the order $10^{10}$ cm$^{-3}$ [10]. From long term observations it was shown that the BLR is compact and at the distances on average of few tens of light days from the central source which is of comparably smaller size scales than the BLR [11]. The BLR gas is photoinozed and lines are being emitted after radiative recombination. Spectropolarimetry of optical broad emission lines have shown that $\varphi$ profiles have "S" shaped profiles around the $\varphi$ of the continuum level, which can be explained if the BLR has disk-like geometry undergoing Keplerian motion and that the light is being scattered by an equatorial scattering region farther away [8]. The characteristic $\varphi$ profiles can be used as an independent way for estimating SMBHs [12] and it was succesfully done for around thirty objects so far [13]. The limitations of this method as well as the polarization sigitures of Type-1 objects were extensively discussed by Savic et al. [14].

In a recent work by Savic et al. [15], it was found that the presence of the SMBBH can influence the profiles of polarization angles which drasticaly deviate than the ones tipically found in Type-1 objects where scattering induced polarization is dominant [8]. In this paper we compare the $\varphi$ profiles arising due to the complex motions such as outflows that could be present in the BLR with the $\varphi$ profiles due to the possible presence of the SMBBH in AGNs. In Sec. 2 we describe the model we used. In Sec. 3 we present obtained results. In Sec. 4 a discussion is presented, followed by a conclusion in Sec. 5.

## 2. Model setup and radiative transfer

We use radiative transfer code *STOKES* [16, 17, 18, 19] for investigating the scattering induced polarization of broad emission lines. The code is based on Monte Carlo algorithm and is capable of numerically solving 3D radiative transfer with kinematics. The basic principle of the code is to generate large number of photons per wavelength bin which would obey a given spectral energy distribution (typically power-law for the continuum and Gaussian or Lorentzian profiles for spectra lines). After the photons leave the defined emitting regions they can be scattered once or many times or absorbed before reaching the observer. The fate of each photon after every scattering event is determined using random numbers. A uniform grid of virtual detectors which surrounds the system save the polarization state of each photon given by Stokes parameters I, Q, U and V . The total intensity, degree of polarization and $\varphi$ for each detector are computed in the end when all photons are collected. The code is publically available and the latest online version is 1.2.[1] As a default output of *STOKES*, we adopt the same convention for orthogonal polarization and parallel polarization as following: $\varphi = 90°$ for parallel or $\varphi = 0°$ for orthogonal polarization [16].

### 2.1 Model geometry

We assume that the SMBHs are at sub-pc scale, but far enough, so the motion of the system can be described using well known equations for two body problem [20]. One of the major assumptions is that each black hole has it's own accretion disk and the corresponding BLRs which are co-planar. The assumption of co-planarity is well justified by numerical simulations which have shown that the angular momentum of the binary aligns with the angular momentum of the inspiraling gas in timescale that is only a fraction of the total evolution time of the binary. The line shapes emitted from these systems can be very complex [21, see for a detailed review].

We used the same model geometry as given by Savic et al. [15]. Each SMBH has the mass of $5 \times 10^7$ Solar mass. Four cases were treated depending on the distance and the shape of the BLRs configuration: **distant**, **contact**, **mixed** and **spiral**. For a single SMBH, we keep the same size of the BLR, but with the SMBH mass of $10^8$ Solar mass. Only for a model with a single SMBH, we allowed vertical outflows in the inner part of the BLR besides the Keplerian motion. An illustration of the model is shown in Fig. 1. We modeled BLR and the scattering region (SR) with flared-disk geometry [16, 17] with half-opening angle of 25 and 30 degrees respecitvely. We adopted the same model parameters as the ones by Savic et al. [15, see Table 1 for full list of parameters]. For the continuum emission, we used a point source approximation with SED given by a power-law $F_c \propto \nu^{-\alpha}$, where $\nu$ is frequency and $\alpha$ is spectral index. We set $\alpha = 2$, which gives constant flux in wavelength space. The scattering region surrounds the central engine with inner and outer radius of 0.1 and 0.5pc respectively. We assume Thomson scattering as a dominant polarizing mechanism in Type-1 objects [5, 22]. Total radial optical depth is 3, which is the upper limit on producing the polarization signal found by spectropolarimetric observations [17].

## 3. Results

We simulated equatorial scattering for four models with SMBBH and one model with SMBH in the central engine. In

---

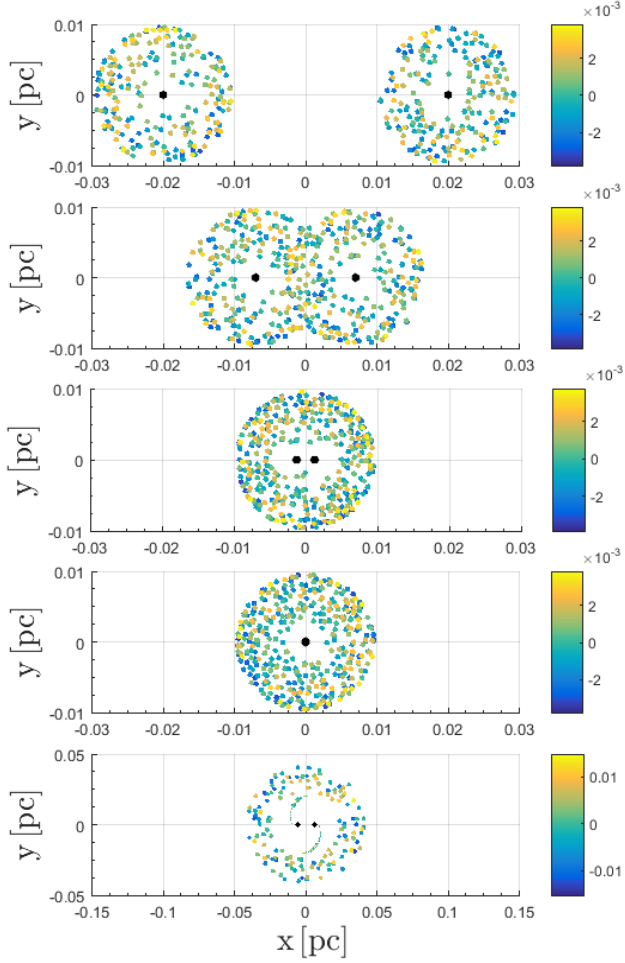

Fig. 1: A sketch depicting the model geometry for four different binary scenarios and one case with a single SMBH. Each BLR clump is denoted by a filled circle with color representing vertical offset along the vertical direction. Black spheres denote the position of each SMBH. From top to bottom: **distant**, **contact**, **mixed**, single SMBH case and **spiral** model. We point out that the velocity field is not denoted, but it was calculated in the same way as by Savic et al. 2018 [15].

Fig. 2 we show the results for $\varphi$ for all binary models compared with the $\varphi$ profile for a single SMBH scenario.

. We chose nearly pole-on viewing inclination angle and 18 degrees azimuthal viewing angle. The nearly pole-on viewing inclination gives the highest amplitude in $\varphi$ change which suits the best for comparative purpose. For single SMBH the system is axis-symmetric. For each of the four binary models, we show simulated $\varphi$ (dashed line), the simulated $\varphi$ for a single SMBH (solid line) and the difference between them (dotted line).

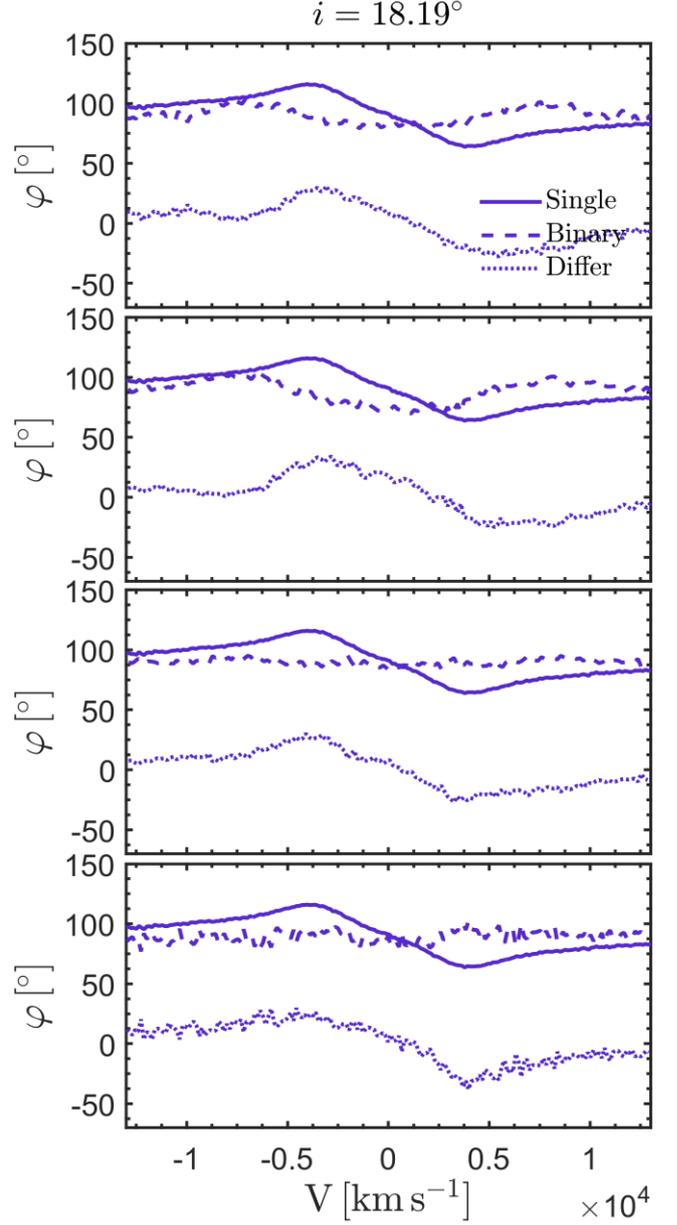

Fig. 2: Polarization angle against velocity for four binary models (dashed line) in comparison with the single SMBH model (solid line). Difference between the two is denoted by dotted line. From top to bottom: **distant**, **contact**, **mixed**, and **spiral**.

Model with a single SMBH shows typical point-symmetric profiles (the function that describes it is odd) with $\varphi$ amplitude in the blue part of the line of around 25 degrees above the continuum level, followed by a drop for the same amount in the red part of the line. Farther in the wings, the profile slowly tends to reach the continuum level. The influence of the outflows only affects the $\varphi$ amplitude by reducing it's value for roughly 5 degrees, since the outflowing velocity is less than one third of the Keplerian velocity in the innermost part of the BLR.

For **distant** model (Fig. 2, top panel), we can see that $\varphi$ profiles are double-peaked and that it is axis-symmetric with respect to the zero-velocity line (described by even function of velocity). The $\varphi$ amplitude is around 20 degrees in the wings, followed by a minimum in the line core. The $\varphi$ reaches the continuum level faster than in the case for one SMBH. The opposite case with $\varphi$ minima in the wings and a maximum in the core is also possible for different azimuthal viewing angles [15, see for detailed results]. The $\varphi$ profile clearly differs from the case with a single SMBH and at some point the difference between them even reaches 40 degrees.

The **contact** model shows very similar $\varphi$ profile as distant model (Fig. 2, top second panel). The $\varphi$ profile is double-peaked with amplitude higher for few degrees than in the previous case. There is a light asymmetry of the $\varphi$ profile around the minimum which is displaced from the center, but it is only due to the finite number of clouds we generated in the model. This case also deviates largely from the familiar profile for a single SMBH. The maximal difference between the two profiles also reaches roughly 50 degrees

The results for the **mixed** model are shown in Fig. 2 (bottom second panel). The $\varphi$ profile is rather flat with very few visible characteristic. The $\varphi$ changes around the continuum level are low (less than 10 degrees). The difference profile is almost the same as the profile for the single SMBH except in the wings where the $\varphi$ profile for mixed has the highest deviations from the continuum level.

The results for the **spiral** model are shown in Fig. 2 (bottom panel). The resulting $\varphi$ profile is complex with a double-peaked feature in the wings with the amplitude of 10 degrees and peak velocity which is close to the orbital velocity of each component of the binary. Closer to the core, there are two minima and one local maximum in at the zero-velocity. The $\varphi$ profile is axis-symmetric, same as the results for distant and contact models. The difference profile is lower in the blue part due to the blue peaks for both models being above the continuum level, while for the red part, this difference is higher since the $\varphi$ profile for a single SMBH reaches minimum below the continuum level.

## Summary and discussion

We compared the results of the simulated $\varphi$ profiles with the $\varphi$ profile for a single SMBH with outflows present in the BLR since the $\varphi$ profiles are very sensitive to geometry and kinematics of the system [5, 14, 23]. We have shown that there is a clear difference in the $\varphi$ profiles between the binary and the single SMBH model, namely in the symmetry of the profiles. Profiles for SMBBHs are axis-symmetric with respect to the zero velocity line, which yields double-peaked profiles. On the contrary, $\varphi$ profiles for one SMBH is point symmetric even with complex motions including outflows. The $\varphi$ amplitude for binary models is less than 20 degrees and the peaks are shifted more towards the wing, which is in agreement with the results by Savic et al. [15]. That was not the case for a single SMBH where $\varphi$ minimum and maximum are closer to the core and with values greater than 20 degrees. Measuring the mass of the binary system proved to be impossible using the AP15 method. Even for the merged model which is the closest to the model with a single SMBH. This is counter-intuitive and reflects how even the low asymmetry in the velocity field can have a huge impact on the $\varphi$ profiles.

## Conclusions

We simulated equatorial scattering for different SMBBH configurations for five simple and comprehensive models using numerical 3D Monte Carlo radiative transfer for scattering induced polarization of the broad emission lines in Type-1 AGNs. From the comparison between the SMBBH and SMBH models, we can conclude the following:

- The $\varphi$ profiles for SMBBH models produce the axis-symmetric profiles which are often double or multi-peaked.

- The $\varphi$ profile for a single SMBH model show point-symmetric profiles even when the additional motions in the BLR are present.

We pointed out that the high quality optical spectropolarimetry of the broad emission lines might play a promising role in the search for the SMBBHs in the future. In the following work, we plan to investigate in details the influence of different motions typically observed in high ionization lines such as C IV and Mg on the polarization profiles and how is that affecting SMBH mass estimates using the polarization position angle.

## Acknowledgements

This work was supported by the Ministry of Education and Science of the Republic of Serbia through the project Astrophysical Spectroscopy of Extragalactic Objects (176001). The author also thanks the French Government and the French Embassy in Serbia for supporting his research.